\documentclass[prd,a4paper,superscriptaddress,nofootinbib,showpacs]{revtex4}
\usepackage{amsmath}
\numberwithin{equation}{section}

\newcommand{\p}{\partial}
\newcommand{\g}{\gamma}

\newcommand{\n}{\nabla}

\begin{document}

\title{1+3 covariant dynamics of scalar perturbations in braneworlds}

\author{Bernard Leong}
\email{cwbl2@mrao.cam.ac.uk}
\affiliation{Astrophysics Group, Cavendish Laboratory, Madingley Road,
Cambridge CB3 OHE, UK}
\author{Peter Dunsby}
\email{peter@vishnu.mth.uct.ac.za}
\affiliation{Department of Mathematics and Applied Mathematics,
University of Cape Town, Rondebosch 7701, South Africa}
\author{Anthony Challinor}
\email{A.D.Challinor@mrao.cam.ac.uk}
\affiliation{Astrophysics Group, Cavendish Laboratory, Madingley Road,
Cambridge CB3 OHE, UK}
\author{Anthony Lasenby}
\email{A.N.Lasenby@mrao.cam.ac.uk}
\affiliation{Astrophysics Group, Cavendish Laboratory, Madingley Road,
Cambridge CB3 OHE, UK}

\date{\today}

\begin{abstract}
We discuss the dynamics of linear, scalar perturbations in an
almost Friedmann-Robertson-Walker braneworld  cosmology of
Randall-Sundrum type II using the 1+3 covariant approach. We derive
a complete set of frame-independent equations for the total matter variables,
and a partial set of equations for the non-local variables which arise
from the projection of the Weyl tensor in the bulk. The latter equations are
incomplete since there is no propagation equation for the non-local
anisotropic stress. We supplement the equations for the total matter
variables with equations for the independent constituents in a cold
dark matter cosmology, and provide solutions in the high and
low-energy radiation-dominated phase under the assumption that the
non-local anisotropic stress vanishes. These solutions reveal the
existence of new modes arising from the two additional non-local degrees
of freedom. Our solutions should prove useful in setting up initial
conditions for numerical codes aimed at exploring the effect of braneworld
corrections on the cosmic microwave background (CMB) power spectrum. As a first
step in this direction, we derive the covariant form of the line of sight
solution for the CMB temperature anisotropies in braneworld cosmologies, and
discuss possible mechanisms by which braneworld effects may remain in the
low-energy universe.
\end{abstract}
\pacs{04.50.+h, 98.80.Cq}
\maketitle

\section{Introduction}

It is understood that Einstein's theory of general relativity is an
effective theory in the low-energy limit of a more general
theory. Recent developments in theoretical physics, particularly in
string theory or M-theory, have led to the idea that gravity is a
higher-dimensional theory which would become effectively
four-dimensional at lower energies.

Braneworlds, which were inspired by string and M-theory, provide simple,
yet plausible, models of how the extra dimensions might affect the
four-dimensional world we inhabit. There is the exciting possibility that
these extra dimensions might reveal themselves through specific
cosmological signatures that survive the transition to the low-energy
universe. It has been suggested that in the context of braneworld models 
the fields that govern the basic interactions in the standard model 
of particle physics are confined to a 3-brane, while the 
gravitational field can propagate
in $3+d$ dimensions (the {\it bulk}). It is not necessarily true
that the extra dimensions are required to be small or even
compact. It was shown recently by Randall and Sundrum~\cite{randall}
for the case $d=1$, that gravity could be localized to a single 3-brane
even when the fifth dimension was infinite. As a result, the Newtonian
potential is recovered on large scales, but with a
leading-order correction on small scales:
\begin{equation}
V(r) = -\frac{GM}{r} \left( 1 + \frac{2l^2}{3r^2} \right)\;,
\end{equation}
where the 5-dimensional cosmological constant $\tilde{\Lambda} \propto
-l^{-2}$. As a result, general relativity is recovered in 4 dimensions
in the static weak-field limit, with a first-order correction which is
believed to be constrained by sub-millimeter experiments at the TeV
level \cite{randall,maartens3}. 

The cosmic microwave background (CMB) currently occupies a central role
in modern cosmology. It is the cleanest cosmological observable, providing
us with a unique record of conditions along our past light cone back
to the epoch of decoupling when the mean free path to Thomson scattering
rose suddenly due to hydrogen recombination. Present
(e.g. BOOMERANG \cite{boomerang1}) and MAXIMA \cite{maxima1}) and
future (e.g. MAP and PLANCK) data on the CMB anisotropies and large-scale
structure provide extensive information on the spectrum and
evolution of cosmological perturbations potentially allowing us to infer the
spectrum of initial perturbations in the early universe and to
determine the standard cosmological parameters to high accuracy. An obvious
question to ask is whether there are any signatures of extra dimensions
which could be imprinted on the cosmic microwave sky.

The aim of this paper is to set up the evolution and constraint equations
for perturbations in a cold dark matter (CDM) brane cosmology,
presenting them in such a way that they can be readily compared with the
standard four-dimensional results, and to provide approximate solutions
in the high and low-energy universe under certain restrictions on how the
bulk reacts on the brane. Our equations are clearly incomplete
since they lack a propagation equation for the non-local
anisotropic stress that arises from projecting the bulk Weyl tensor onto
the brane, and our solutions are only valid under the neglect of this stress.
However, our presentation is such that we can easily include effective
four-dimensional propagation equations for the non-local stress should
such equations arise from a study of the full bulk perturbations.
The lack of a four-dimensional propagation equation for the non-local
stress means that it is currently not possible to obtain general results
for the anisotropy of the CMB in braneworld models. Such a calculation
would require solving the full five-dimensional perturbation equations which
is non-trivial since the equations can only be reduced to two-dimensional
partial differential equations on Fourier transforming the 3-space dependence.
Only qualitative results are currently known, obtained with either the
standard metric-based (gauge-invariant) approach
\cite{anchordoqui,bridgman,copeland,deruelle,gorbunov,kodama,kodama2,koyama,langlois,langlois2,langlois3,langlois4,mukohyama,vandebruck,vandebruck2,vandebruck3}, or with 1+3 covariant methods 
\cite{gordon,maartens,maartens2}.

In order to make this paper self-contained, we begin by giving
a brief overview of the 1+3 covariant approach to
cosmology and define the key variables we use to characterize
the perturbations in Sec.~\ref{sec:cov}. After a short review
of how this approach can be used to describe the general dynamics
of Randall-Sundrum braneworlds, in Sec.~\ref{sec:equations}
we present a complete set of frame-independent, linear equations
describing the evolution of the total matter variables and the non-local
energy and momentum densities for scalar perturbations in an almost-FRW
universe (with arbitrary spatial curvature).
Many of these equations, which employ only covariantly defined,
gauge-invariant variables, have simple Newtonian analogues
\cite{ellis89}, and their physical meanings are considerably more
transparent than those that underlie more standard metric-based
approaches. In Sec.~\ref{sec:radiation} we derive
analytic solutions for the scalar modes in the high and low-energy
radiation-dominated universe neglecting the non-local anisotropic
stress. In principle, these solutions could be used in a
phenomenological manner to generate more general solutions
which include non-local anisotropic stress, using
Green's method and an ansatz for the non-local stress. Such a study
will be presented in a subsequent paper~\cite{leong2}, along with a numerical
calculation of the CMB power spectrum employing this phenomenology.
As a first step, we derive the covariant line of sight solution for the
temperature anisotropies from scalar modes in braneworld models in
Sec.~\ref{sec:aniso} and present some comments on how higher dimensional
effects may remain in the low-energy universe and thus imprint on the microwave
sky. An appendix presents the scalar perturbation equations in the
matter energy frame during radiation domination.

\section{1+3 covariant decomposition} \label{sec:cov}

Throughout this paper we adopt the metric signature $(-+++)$. Our
conventions for the Riemann and Ricci tensor are fixed by
$[\nabla_a, \nabla_b] U_c = R_{abc}{}^{d} U_d$ and $R_{ab} =
R_{acb}{}^{c}$. Lowercase latin indices $a\ldots b$ are used to denote
the standard 4-dimensional (1+3) spacetime whereas uppercase $A\ldots B$
and the tilde of any physical quantity are used to denote 5-dimensional
(1+4) spacetime (of the braneworld).
Round (square) bracket denote symmetrization (antisymmetrization)
on the enclosed indices.
We use units with $\hbar=c=1$, so the 4-dimensional gravitational
constant is related to the 4-dimensional Planck mass via $G=M_P{}^{-2}$.

We start by choosing a 4-velocity $u^a$. This must be physically
defined in such a way that if the universe is exactly FRW, the velocity
reduces to that of the fundamental observers to ensure gauge-invariance of
the approach. From the 4-velocity $u^a$,
we construct a projection tensor $h_{ab}$ into the space perpendicular
to $u^a$ (the instantaneous rest space of observers whose
4-velocity is $u^a$):
\begin{equation} \label{e:projection1}
h_{ab} \equiv g_{ab} + u_a u_b\;,
\end{equation}
where $g_{ab}$ is the metric of the spacetime. The operation of projecting
a tensor fully with $h_{ab}$, symmetrizing, and removing the trace on every
index (to return the projected-symmetric-trace-free or PSTF part) is
denoted by angle brackets, i.e.\ $T_{\langle ab\ldots c\rangle}$.

The symmetric tensor $h_{ab}$ is used to define a projected (spatial)
covariant derviative $D^a$ which when acting on a tensor
$T^{b\ldots c}{}_{d\ldots e}$ returns a tensor that is orthogonal to $u^a$ on
every index, 
\begin{equation} \label{e:projection2}
D^a T^{b\ldots c}{}_{d\ldots e}
\equiv h^{a}{}_{p} h^{b}{}_{r}
\ldots  h^{c}{}_{s} h^{t}{}_{d} \ldots  h^{u}{}_{e} \n^p T^{r..s}{}_{t..u}\;,
\end{equation}
where $\n^a$ denotes the usual covariant derivative.

The covariant derivative of the 4-velocity can be decomposed
as follows:
\begin{equation} \label{e:covariant1}
\n_a u_b = \omega_{ab} + \sigma_{ab} + \frac{1}{3} \Theta h_{ab} - u_a
A_b\;,
\end{equation}
where $\omega_{ab}$ is the vorticity which satisfies $u^a
\omega_{ab}=0$, $\sigma_{ab}=\sigma_{\langle ab \rangle}$
is the shear which is PSTF, $\Theta \equiv \n^{a} u_a = 3H$
measures the volume expansion rate (where $H$ is the local Hubble
parameter), and $A_a \equiv u^b \n_b u_a$ is the acceleration.

Gauge-invariant quantities can be constructed from scalar variables
by taking their projected gradients. Such quantities vanish in the
FRW limit by construction. The comoving fractional projected gradient
of the density field $\rho^{(i)}$ of a species $i$ (for example, photons) is
one important example of this construction:
\begin{equation} \label{e:sg1}
\Delta_a^{(i)} \equiv \frac{a}{\rho^{(i)}} D_a \rho^{(i)} \;,
\end{equation}
where $a$ is a locally defined scale factor satisfying
\begin{equation} \label{e:hubble}
\dot{a} \equiv u^b \n_b a = Ha\;,
\end{equation}
which is included to remove the effect of the expansion on the projected
gradients. Another important vector variable is the 
comoving projected gradient of the expansion,
\begin{equation} \label{e:sg2}
{\cal Z}_a \equiv a D_a \Theta\;,
\end{equation}
which provides a measure of the inhomogeneity of the expansion.

The matter stress-energy tensor $T_{ab}$ can be decomposed irreducibly
with respect to $u^a$ as follows:
\begin{equation} \label{e:emequation1}
T_{ab} \equiv \rho u_a u_b + 2 u_{(a}q_{b)} + P h_{ab} + \pi_{ab}\;,
\end{equation}
where $\rho \equiv  T_{ab} u^a u^b$ is the energy density measured
by an observer moving with 4-velocity $u^a$,
$q_a \equiv - h^{b}{}_{a} T_{bc} u^c$ is the energy flux or momentum density
(orthogonal to $u^a$), $P \equiv h_{ab} T^{ab}/3$ is the isotropic
pressure, and the PSTF tensor $\pi_{ab} \equiv T_{\langle a b\rangle}$ is
the anisotropic stress.

The remaining gauge-invariant variables are formed
from the Weyl tensor $C_{abcd}$ which vanishes in an exact FRW universe
because these models are conformally flat. The ten degrees of freedom in the
4-dimensional Weyl tensor can be encoded in two PSTF tensors: the electric
and magnetic parts defined respectively as
\begin{align}
E_{ab} &= C_{abcd} u^b u^d\;,  \label{e:eweyl} \\
H_{ab} &= \frac{1}{2} C_{acst} u^{c} \eta^{st}{}_{bd} u^d\;,
\label{e:bweyl}
\end{align}
where $\eta_{abcd}$ is the 4-dimensional covariant permutation
tensor.

\section{Field equations of the braneworld}

In a recent paper, Maartens~\cite{maartens} introduced a formalism for
describing the non-linear, intrinsic dynamics of the brane in Randall-Sundrum
type II braneworld models in the form of bulk corrections to the 1+3 covariant
propagation and constraint equations of general relativity.
This approach is well suited to identifying the geometric and physical
properties which determine homogeneity and anisotropy on the brane, and serves
as a basis for developing a gauge-invariant description of
cosmological perturbations in these models.

An important distinction between braneworlds and general relativity
is that the set of 1+3 dynamical equations does not close on the brane.
This is because there is no propagation equation for the non-local effective
anistropic stress that arises from projecting the bulk Weyl tensor onto the
brane. The physical implication is that the initial value problem cannot be
solved by brane-bound observers. The non-local Weyl variables
enter crucially into the dynamics (for example,
the Raychaudhuri equation) of the intrinsic geometry of the brane.
Consequently, the existence of these non-local effects leads to the violation
of several important results in theoretical cosmology, such as the connection
between isotropy of the CMB and the Robertson-Walker geometry.

The field equations induced on the brane are derived by Shiromizu et
al~\cite{shiromizu} using the Gauss-Codazzi equations, together with
the Israel junction conditions and $Z_2$ symmetry.
The standard Einstein equation is modified with new terms
carrying the bulk effects on the brane:
\begin{equation} \label{e:einstein1}
G_{ab} = - \Lambda g_{ab} + \kappa^2 T_{ab} + \tilde{\kappa}^4 {\cal S}_{ab}
- {\cal E}_{ab}\;,
\end{equation}
where $\kappa^2=8 \pi/M_p{}^2$. The energy scales are related to each
other via
\begin{align}
\label{e:constant1} \lambda &= 6 \frac{\kappa^2}{\tilde{\kappa}^4}\;, \\
\label{e:constant} \Lambda & = \frac{1}{2}
\tilde{\kappa}^2 \left(\tilde{\Lambda} + \frac{1}{6} \tilde{\kappa}^2
\lambda^2 \right)\;,
\end{align}
where $\tilde{\Lambda}$ is the cosmological constant in the bulk and
$\lambda$ is the tension of the brane. The bulk corrections to the
Einstein equations on the brane are made up of 
two parts: (i) the matter fields which contribute local quadratic
energy-momentum corrections via the symmetric tensor ${\cal S}_{ab}$;
and (ii) the non-local effects from the free gravitational field in the
bulk transmitted by the (symmetric) projection ${\cal E}_{ab}$ of the bulk
Weyl tensor. The matter corrections are given by
\begin{equation} \label{e:emtensor2}
{\cal S}_{ab} = \frac{1}{12} T_{c}{}^{c} T_{ab} - \frac{1}{4} T_{ac}
T^{c}{}_b + \frac{1}{24} g_{ab} [3 T_{cd} T^{cd} - (T_{c}{}^{c})^2]\;.
\end{equation}
We note that the local part of the bulk gravitational field is the
five dimensional Einstein tensor $\tilde{G}_{AB}$, which is determined
by the bulk field equations. Consequently, ${\cal E}_{ab}$
transmits non-local gravitational degrees of freedom from the
bulk to the brane that includes both tidal (or Coulomb),
gravito-magnetic, and transverse traceless (gravitational wave)
effects.

The bulk corrections can all be consolidated into an effective total
energy density, pressure, anisotropic stress and energy flux. The
modified Einstein equations take the standard Einstein form with a
re-defined energy-momentum tensor:
\begin{equation} \label{e:einstein2}
G_{ab} = - \Lambda g_{ab} + \kappa^2 T^{\text{tot}}_{ab}\;,
\end{equation}
where
\begin{equation} \label{e:emtensor3}
T^{\text{tot}}_{ab} = T_{ab} + \frac{\tilde{\kappa}^4}{\kappa^2}
{\cal S}_{ab} - \frac{1}{\kappa^2} {\cal E}_{ab}\;.
\end{equation}
Decomposing ${\cal E}_{ab}$ irreducibly with respect to $u^a$ by analogy
with Eq.~(\ref{e:emequation1})~\cite{gordon,maartens,maartens2},
\begin{equation}
{\cal E}_{ab} = - \left(\frac{\tilde{\kappa}}{\kappa}\right)^4
\left( {\cal U} u_a u_b + 2 u_{(a}{\cal Q}_{b)} + \frac{\cal U}{3} h_{ab} +
{\cal P}_{ab} \right), 
\end{equation}
(the prefactor is included to make e.g. ${\cal U}$ have dimensions of energy
density), it follows that the total density, pressure, energy flux and
anisotropic pressure are given as follows:
\begin{align}
\rho^{\text{tot}} &= \rho + \frac{\tilde{\kappa}^4}{\kappa^6}
\left[\frac{\kappa^4}{24} (2 \rho^2 - 3 
\pi^{ab} \pi_{ab}) + {\cal U} \right]\;, \\
\label{e:pressure1}
P^{\text{tot}} &= P  + \frac{\tilde{\kappa}^4}{\kappa^6}
\left[\frac{\kappa^4}{24} \left(2 \rho^2 + 4 P \rho + \pi^{ab} \pi_{ab} - 4
q_{a} q^{a} \right) + \frac{1}{3} {\cal U} \right]\;, \\
\label{e:flux}
q^{\text{tot}}_{a} &= q_a + \frac{\tilde{\kappa}^4}{\kappa^6}
\left[\frac{\kappa^4}{24} (4 \rho q_a - 6\pi_{ab} q^{b}) + {\cal Q}_a \right]\;,
\\
\label{e:pressure2}
\pi^{\text{tot}}_{ab} &= \pi_{ab} +  \frac{\tilde{\kappa}^4}{\kappa^6}
\left[\frac{\kappa^4}{12} \left\{ -(\rho + 3P) \pi_{ab} -3 \pi_{c
\langle a}\pi_{b\rangle}{}^{c} + 3q_{\langle a}q_{b\rangle} \right\} +
{\cal P}_{ab} \right]\;.
\end{align}

For the braneworld case, it is useful to introduce an additional
dimensionless gradient which describes inhomogeneity in the non-local energy
density ${\cal U}$:
\begin{equation}
\label{e:nonlocal0}
\Upsilon_a \equiv \frac{a}{\rho} D_a {\cal U}\;.
\end{equation}

The Gauss-Codazzi scalar equation for the 3-curvature defined by ${\cal
R}$ is given by
\begin{equation} \label{e:curvature1}
{\cal R} = 2 \kappa^2 \rho + \frac{1}{6} \tilde{\kappa}^4 \rho^2 + 2
\left(\frac{\tilde{\kappa}}{\kappa} \right)^4 {\cal U} - \frac{2}{3} \Theta^2
+ 2 \Lambda\;,
\end{equation}
where
\begin{equation} \label{e:curvature2}
{\cal R} \equiv ~^{(3)}R = h^{ab}~^{(3)}R_{ab}
\end{equation}
with ${}^{(3)} R_{ab}$ the intrinsic curvature of the surfaces orthogonal
to $u^a$\footnote{If the
vorticity is non-vanishing flow-orthogonal hypersurfaces will not exist, and
${\cal R}$ cannot be interpreted as the spatial curvature scalar.}.
In FRW models the Gauss-Codazzi constraint reduces to the modified
Friedmann equation
\begin{equation}\label{e:friedmann1}
H^2 + \frac{K}{a^2} = \frac{1}{3}\kappa^2 \rho + \frac{1}{3} \Lambda
+ \frac{1}{36} \tilde{\kappa}^4 \rho^2 + \frac{1}{3}\left(\frac{\tilde{\kappa}}
{\kappa}\right)^4 {\cal U},
\end{equation}
where the 3-curvature scalar is ${\cal R}=6K/a^2$.
In non-flat models ($K \neq 0$) ${\cal R}$ is not gauge-invariant since it
does not vanish in the FRW limit. However, the comoving projected gradient
\begin{equation} \label{e:curvature3}
\eta_b \equiv \frac{a}{2} D_b {\cal R}\;
\end{equation} 
is a gauge-invariant measure of inhomogeneity in the intrinsic three
curvature of the hypersurfaces orthogonal to $u^a$.

\section{Linearised scalar perturbation equations for the
total matter variables} \label{sec:equations}
\subsection{Local and non-local conservation equations}

Based on the form of the bulk energy-momentum tensor and $Z_2$
symmetry, the brane energy-momentum tensor is still covariantly conserved:
\begin{equation} \label{e:em1}
\n^b T_{ab} = 0\;.
\end{equation}
The contracted Bianchi identities on the brane ensure conservation of the total
energy-momentum tensor, which combined with conservation of the
matter tensor gives
\begin{equation} \label{e:bianchi1}
\n^{a} {\cal E}_{ab} = \tilde{\kappa}^4 \n^{a} {\cal S}_{ab}\;.
\end{equation}
The longitudinal part of ${\cal E}_{ab}$ is sourced by quadratic
energy-momentum terms including spatial gradients and time derivatives.
As a result any evolution and inhomogeneity in the matter fields
would generate non-local Coulomb-like gravitational effects
in the bulk which back react on the brane. The conservation
equation~(\ref{e:em1}) implies evolution equations for the energy and
momentum densities, and these are unchanged from their general relativistic
form. To linear order in an almost-FRW brane cosmology we have
\begin{equation} \label{e:em2}
\dot{\rho} + \Theta (\rho + P) + D^a q_a = 0 \;,
\end{equation}
and
\begin{equation} \label{e:em3}
\dot{q}_a + \frac{4}{3} \Theta q_{a}  + (\rho +P )A_a + D_a P + D^b
\pi_{ab} = 0\;.
\end{equation}
The linearised propagation equations for ${\cal U}$ and ${\cal Q}$ follow from
Eq.~(\ref{e:bianchi1}) (see Ref.~\cite{maartens}):
\begin{equation} \label{e:nonlocal1}
\dot{{\cal U}} + \frac{4}{3} \Theta {\cal U} + D^a {\cal Q}_a =0 \;,
\end{equation}
and
\begin{equation} \label{e:nonlocal2}
\dot{{\cal Q}_{a}} +\frac{4}{3} \Theta {\cal Q}_{a} +
\frac{1}{3} D_a {\cal U} + D^{b} {\cal P}_{ab} + \frac{4}{3} {\cal U}A_a
= \frac{\kappa^4}{12} (\rho + P) \left(-2 D_a \rho + 3 D^b
\pi_{ab} + 2 \Theta q_a \right)\;.
\end{equation}

Taking the projected derivative of Eq.~(\ref{e:em2}) we obtain the
propagation equation for $\Delta_a$ at linear order:
\begin{equation}
\rho \dot{\Delta}_a + (\rho +P)({\cal Z}_a + a \Theta A_a) + a
D_a D^b q_b + a \Theta D_a P - \Theta P \Delta_a = 0.
\end{equation}
From equation $\eqref{e:nonlocal1}$, we obtain the evolution equation
of the spatial gradient of the non-local energy density: 
\begin{equation}
\dot{\Upsilon}_a = \left(\frac{P}{\rho} - \frac{1}{3} \right) \Theta
\Upsilon_a - \frac{4}{3} \frac{{\cal U}}{\rho}({\cal Z}_a + a\Theta A_a)
- \frac{a}{\rho} D_a D^b {\cal Q}_b.
\end{equation}

From the propagation equations for ${\cal U}$ and ${\cal Q}$ it can be seen
that the energy of the
projected Weyl fluid is conserved while the momentum is not conserved;
rather it is driven by the matter source terms on the right of
Eq.~(\ref{e:bianchi1}). Note that no propagation equation for ${\cal P}_{ab}$
is implied so the set of equations will not close.

\subsection{Propagation and constraint equations}

In this section we give the linearised gravito-magnetic and
gravito-electric propagation and constraint scalar equations on the brane,
which follow
from the Bianchi identities, and the equations
for the kinematic variables $\sigma_{ab}$, and $\Theta$ and its gradient
${\cal Z}_a$ which follow from the Ricci identity.

For scalar perturbations, the magnetic part of the Weyl tensor
$H_{ab}$ and the vorticity tensor $\omega_{ab}$ vanish identically.
The electric part of the Weyl tensor $E_{ab}$ and the
shear $\sigma_{ab}$ need not vanish. The non-vanishing variables
satisfy the following propagation and constraint equations on the brane:
\begin{enumerate}
\item Gravito-electric propagation:
\begin{equation} \label{e:propagation1}
\begin{split}
&\dot{E}_{ab} + \Theta E_{ab} + \frac{1}{2} \kappa^2 (\rho + P)
\sigma_{ab} + \frac{1}{2} \kappa^2 D_{\langle a}q_{b \rangle} +
\frac{1}{6} \kappa^2 \Theta \pi_{ab} + \frac{1}{2} \kappa^2 \dot{\pi}_{ab}\\
&= \frac{1}{72} \left(\frac{\tilde{\kappa}}{\kappa}\right)^4 [\kappa^4 \{-6\rho
(\rho + P) \sigma_{ab} + 3 (\dot{\rho} + 3\dot{P}) \pi_{ab}
+ 3 (\rho + 3 P) \dot{\pi}_{ab}   \\
&- 6\rho D_{\langle a} q_{b\rangle}
+ \Theta [\rho + 3 P] \pi_{ab}\} - 48 {\cal U} \sigma_{ab} - 36
\dot{{\cal P}}_{ab} - 36 D_{\langle a}{\cal
Q}_{b\rangle} - 12 \Theta {\cal P}_{ab}]\; ;
\end{split}
\end{equation}
\item Shear propagation:
\begin{equation} \label{e:propagation2}
\dot{\sigma}_{ab} + \frac{2}{3} \Theta \sigma_{ab} +
E_{ab} -
\frac{1}{2} \kappa^2 \pi_{ab} - D_{\langle a} A_{b\rangle} = \frac{1}{24}
\left(\frac{\tilde{\kappa}}{\kappa}\right)^4 \{ \kappa^4 [ - (\rho + 3P)
\pi_{ab} ] + 12 {\cal P}_{ab}\}\; ;
\end{equation}
\item Shear constraint:
\begin{equation} \label{e:constraint1}
D^{b} \sigma_{ab} - \frac{2}{3} D_{a} \Theta + \kappa^2 q_a =
-\frac{1}{6} \left(\frac{\tilde{\kappa}}{\kappa}\right)^4 (\kappa^4 \rho
q_a + 6 {\cal Q}_{a})\; ;
\end{equation}
\item Gravito-electric divergence:
\begin{equation} \label{e:constraint2}
\begin{split}
D^{b} E_{ab} + \frac{1}{2} \kappa^2 D^{b} \pi_{ab} - \frac{1}{3}
\kappa^2 D_{a} \rho + \frac{1}{3} \kappa^2 \Theta q_a &=
\frac{1}{48}\left(\frac{\tilde{\kappa}}{\kappa}\right)^4
\bigg[\kappa^4 \left(- \frac{8}{3} \rho \Theta q_a + 2 (\rho + 3P)D^{b}\pi_{ab} +
\frac{8}{3} \rho D_a \rho \right) \\
& \quad + 16 D_a {\cal U}  - 16 \Theta {\cal Q}_a - 24 D^b
{\cal P}_{ab} \bigg]\; ;
\end{split}
\end{equation}
\item Modified Raychaudhuri equation:
\begin{equation} \label{e:raychaudhuri}
\dot{\Theta}= -\frac{1}{3}\Theta^2 - \frac{1}{2}\kappa^2(\rho+3P)
+ \Lambda - \frac{1}{12}\left(\frac{\tilde{\kappa}}{\kappa}\right)^4 [\kappa^4 \rho
(2\rho+3P) + 12 \mathcal{U}] + D^a A_a\; ; 
\end{equation}
\item Propagation equation for the comoving expansion gradient
${\cal Z}_a$ which follows from Eq.~$\eqref{e:raychaudhuri}$:
\begin{equation} \label{e:propagation3}
\dot{{\cal Z}}_a + \frac{2}{3} \Theta \mathcal{Z}_{a} - a \dot{\Theta} A_a
+ \frac{\kappa^2}{2} aD_a(\rho+3P) - aD_a D^b A_b
= -\frac{1}{12} \left(\frac{\tilde{\kappa}}{\kappa}\right)^4 \{
\kappa^4 aD_a[\rho(2\rho+3P)] + 12 a D_a {\cal U} \} \;.
\end{equation}
\end{enumerate}

The spatial gradient of the 3-curvature scalar is an auxiliary
variable. It can be related to the other gauge-invariant variables
using Eqs.~$\eqref{e:curvature1}$ and $\eqref{e:curvature3}$:
\begin{equation} \label{e:curconstraint}
\eta_a = \kappa^2 \rho \Delta_a + \frac{1}{6}\tilde{\kappa}^4 \rho^2 \Delta_a
+ \left(\frac{\tilde{\kappa}}{\kappa} \right)^4 \rho \Upsilon_a - \frac{2}{3}
\Theta {\cal Z}_a\;.
\end{equation}

Taking the time derivative of Eq.~$\eqref{e:curconstraint}$,
commuting the spatial and temporal
derivatives, and then making use of Eqs.~$\eqref{e:raychaudhuri}$ and
$\eqref{e:propagation3}$, we obtain the evolution of the spatial
gradient of the 3-curvature scalar:
\begin{equation} \label{e:evolution}
\dot{\eta}_a + \frac{2}{3} \Theta \eta_a + \frac{1}{3} {\cal R}
({\cal Z}_a + a \Theta A_a) + \frac{2}{3}\Theta a D_a D^b A_b
= -\left(\kappa^2 + \frac{1}{6} \tilde{\kappa}^4 \rho\right) a D_a D^b q_b
- \left(\frac{\tilde{\kappa}}{\kappa}\right)^4 a D_a D^b {\cal Q}_b.
\end{equation}
In general relativity, propagating $\eta_a$ is a useful device to avoid
numerical instability problems when solving for isocurvature modes in a
zero acceleration frame (such as the rest-frame of the CDM)~\cite{lewis}.

\section{Cosmological scalar perturbations in the braneworld}
\label{sec:scalar}

The tensor-valued, partial differential equations presented in the earlier
sections can be reduced to scalar-valued, ordinary differential equations
by expanding in an appropriate complete set of eigentensors. For scalar
perturbations all gauge-invariant tensors can be constructed from
derivatives of scalar functions. It is thus natural to expand in STF tensors
derived from the scalar eigenfunctions $Q^{(k)}$
of the projected Laplacian:
\begin{equation} \label{e:helmholtz}
D^2 Q^{(k)} = - \frac{k^2}{a^2} Q^{(k)}\;,
\end{equation}
satisfying $\dot{Q}^{(k)} = O(1)$\footnote{The notation $O(n)$ is short
for $O(\epsilon^n)$ where $\epsilon$ is some dimensionless quantity
characterising the departure from FRW symmetry.}.
We adopt the following harmonic
expansions of the gauge-invariant variables:
\begin{equation} \label{e:harmonics1}
\begin{split}
\Delta^{(i)}_a = \sum_k k \Delta_k^{(i)} Q^{(k)}_a\;,
\quad&\quad {\cal Z}_a = \sum_k \frac{k^2}{a} {\cal Z}_k
Q^{(k)}_a\;, \\
q^{(i)}_a = \rho^{(i)} \sum_k q_k^{(i)} Q_a^{(k)}\;,
\quad&\quad \pi^{(i)}_{ab} =  \rho^{(i)} \sum_k \pi_k^{(i)}
Q_{ab}^{(k)}\;, \\
E_{ab} = \sum_k \frac{k^2}{a^2} \Phi_k Q_{ab}^{(k)}\;,
\quad&\quad \sigma_{ab} =   \sum_k \frac{k}{a} \sigma_k Q_{ab}^{(k)}\;,
\\ v^{(i)}_{a} =  \sum_k v^{(i)}_k Q_a^{(k)}\;,  \quad&\quad A_a =  \sum_k
\frac{k}{a} A_k Q_a^{(k)}\;.
\end{split}
\end{equation}
Here $v^{(i)}_a$ is the 3-velocity of species $i$ relative to $u^a$; for the
CDM model considered here we shall make use of $v^{(i)}_a$ for baryons $b$
and CDM $c$. For photons $\gamma$ and neutrinos $\nu$ we continue to work with
the momentum densities which are related to the peculiar velocity of the
energy frame for that species by e.g.\ $q^{(\gamma)}_a = (4/3)\rho^{(\gamma)}
v^{(\gamma)}_a$ in linear theory.
The scalar expansion coefficients, such as $\Delta_k^{(i)}$ are
first-order gauge-invariant variables satisfying  e.g.\ $D^a
\Delta_k^{(i)} = O(2)$. Projected vectors $Q_a^{(k)}$ and STF
tensors $Q_{ab}^{(k)}$ are defined by~\cite{dunsby}:
\begin{equation} \label{e:scalarid1}
\begin{split}
Q_a^{(k)} &= - \frac{a}{k} D_a Q^{(k)}\;, \\
Q_{ab}^{(k)} &= \frac{a^2}{k^2} D_{\langle a} D_{b\rangle }Q^{(k)}\;,
\end{split}
\end{equation}
so that
\begin{equation}
D^b Q_{ab}^{(k)} = \frac{2}{3} \left(\frac{k}{a} \right) \left(1 -
\frac{3K}{k^2} \right) Q_a^{(k)}\;.
\end{equation}
We expand the non-local perturbation variables in scalar harmonics in the
following manner:
\begin{equation} \label{e:harmonics2}
\begin{split}
\Upsilon_a &= \sum_k k \Upsilon_k Q_a^{(k)}\;, \\
{\cal Q}_a &= \sum_k \rho {\cal Q}_k Q_a^{(k)}\;, \\
{\cal P}_{ab} &= \sum_k \rho {\cal P}_k Q_{ab}^{(k)}\;.
\end{split}
\end{equation}
In addition, we can expand the projected gradient of the 3-curvature term:
\begin{equation}
\eta_a = \sum_k 2 \left(\frac{k^3}{a^2} \right) \left(1 - \frac{3K}{k^2}
\right) \eta_k Q_{a}^{(k)} \;. 
\end{equation}
The form of this expansion is chosen so that if we adopt the energy
frame (where $q_a=0$) the variable $\eta_k$ coincides with the curvature
perturbation usually employed in gauge-invariant calculations.

\subsection{Scalar equations on the brane}

It is now straightforward to expand the 1+3 covariant propagation and
constraint equations in scalar harmonics. We shall consider the
CDM model where the particle species are baryons (including
electrons), which we model as an ideal fluid with pressure $p^{(b)}$ and
peculiar velocity $v^{(b)}_a$, cold dark matter, which has vanishing
pressure and peculiar velocity $v^{(c)}_a$, and photons and (massless)
neutrinos which require a kinetic theory description.
We neglect photon polarization,
although this can easily be included in the 1+3 covariant
framework~\cite{challinor3}. Also, we assume that the entropy perturbations
are negligible for the baryons, so that $D_a P^{(b)} = c_s^2 D_a \rho^{(b)}$
where $c_s^2$ is the adiabatic sound speed. 
A complete set of 1+3 perturbation
equations for the general relativistic model can be found
in~\cite{challinor2}. We extend these equations to braneworld models here.

In the following, perturbations in the total matter variables are related to
those in the individual components by
\begin{equation}
\rho \Delta_k = \sum_i \rho^{(i)} \Delta^{(i)}_k, \quad
\rho q_k = \sum_i \rho^{(i)} q^{(i)}_k, \quad
\rho \pi_k = \sum_k \rho^{(i)} \pi^{(i)}_k, 
\end{equation}
where $q^{(b)}_k=(1+P^{(b)}/\rho^{(b)})v^{(b)}_k$, $q^{(c)}_k = v^{(c)}_k$, and
$\pi^{(i)}_k$ vanishes for baryons and CDM. Similarly, the total density
and pressure are obtained by summing over components, e.g.\
$P = \sum_i P^{(i)}$. It is also convenient to write $P=(\gamma-1)\rho$, but
$\gamma$ should not be assumed constant (in space or time).

We begin with the equation for the gravito-electric field:
\begin{equation} \label{e:propagation1a}
\begin{split}
&\left(\frac{k}{a}\right)^2 \left(\dot{\Phi}_k + \frac{1}{3} \Theta \Phi_k \right) +
\frac{1}{2} \frac{k}{a} \kappa^2 \rho (\g \sigma_k - q_k) +
\frac{1}{6} \kappa^2 \rho \Theta (1-3\gamma) \pi_k + \frac{1}{2} \kappa^2
\rho \dot{\pi}_k  \\
&= \frac{1}{72} \left(\frac{\tilde{\kappa}}{\kappa}\right)^4 \bigg\{ -\kappa^4
\bigg[6 \left(\frac{k}{a} \right) \rho^2 (\g \sigma_k -q_k)- 3(\dot{\rho} + 3\dot{P})
\rho \pi_k - 3 (3 \g -2) \rho (\rho \dot{\pi}_k + \dot{\rho} \pi_k)
- (3\g -2) \rho^2 \Theta \pi_k \bigg] \\ & - 12 \left(\frac{k}{a} \right)(4 {\cal U}
\sigma_k - 3 \rho \mathcal{Q}_k) - 36
(\dot{\rho} {\cal P}_k + \rho \dot{{\cal P}}_k) - 12 \rho \Theta
{\cal P}_k \bigg\}\;.
\end{split}
\end{equation}
We have written this equation in such a form that every term is
manifestly frame-independent.
The shear propagation equation is
\begin{equation} \label{e:propagation2a}
\frac{k}{a} \left(\dot{\sigma}_k + \frac{1}{3} \Theta \sigma_k \right) +
\left(\frac{k}{a}\right)^2 (\Phi_k + A_k) - \frac{\kappa^2}{2} \rho \pi_k =
\frac{1}{24}\left(\frac{\tilde{\kappa}}{\kappa}\right)^4 [-(3 \g - 2) \kappa^4
\rho^2 \pi_k + 12 \rho {\cal P}_k]\;.
\end{equation}
The shear constraint is given by
\begin{equation} \label{e:constraint1a}
\kappa^2 \rho q_k - \frac{2}{3} \left(\frac{k}{a} \right)^2 \left[{\cal Z}_k - \left(1
- \frac{3K}{k^2} \right) \sigma_k \right] = - \frac{1}{6}
  \left(\frac{\tilde{\kappa}}{\kappa}\right)^4 (\kappa^4 \rho^2 q_k + 6\rho
  {\cal Q}_k).
\end{equation}
The gravito-electric divergence is
\begin{equation} \label{e:constraint2a}
\begin{split}
& 2 \left(\frac{k}{a} \right)^3 \left(1 - \frac{3K}{k^2} \right) \Phi_k -
\kappa^2 \rho \left(\frac{k}{a} \right) \left[\Delta_k - \left(1 -
\frac{3K}{k^2} \right) \pi_k \right] + \kappa^2 \Theta \rho q_k \\
&= \frac{1}{16} \left(\frac{\tilde{\kappa}}{\kappa}\right)^4
\bigg\{\kappa^4 \left[-\frac{8}{3} \rho^2 \Theta q_k + \frac{4}{3} (3\g -2)
\rho^2 \left(1 - \frac{3K}{k^2}\right) \frac{k}{a} \pi_k  + \frac{8}{3} \frac{k}{a}
\rho^2 \Delta_k \right]  \\
& + 16  \frac{k}{a} \rho \Upsilon_k - 16 \Theta \rho {\cal Q}_k - 16 \rho
\left(\frac{k}{a} \right) \left(1 -
\frac{3K}{k^2} \right) {\cal P}_k \bigg\} \;.
\end{split}
\end{equation}
The  propagation equation for the comoving expansion gradient
${\cal Z}_a$ is given by
\begin{equation} \label{e:propagation3b}
\begin{split}
&\dot{{\cal Z}}_k + \frac{1}{3} \Theta \mathcal{Z}_k - \frac{a}{k}
\dot{\Theta} A_k + \frac{k}{a} A_k + \frac{\kappa^2}{2} \frac{a}{k}
\left[2 (\rho^{(\g)}
\Delta^{(\g)}_k + \rho^{(\nu)} \Delta^{(\nu)}_k ) + (1 + 3 c_s^2)
\rho^{(b)}
\Delta^{(b)}_k + \rho^{(c)} \Delta^{(c)}_k  \right] \\
&= -\frac{1}{12}  \left(\frac{\tilde{\kappa}}{\kappa}\right)^4 \frac{a}{k}
\left\{\kappa^4[(2\rho + 3P)\rho\Delta_k + \rho(3\rho^{(\g)}\Delta^{(\g)}_k
+3\rho^{(\nu)}\Delta^{(\nu)}_k + (2+3c_s^2)\rho^{(b)}\Delta_k^{(b)} +
2 \rho^{(c)} \Delta^{(c)}_k)] + 12 \rho \Upsilon_k \right\}\;.
\end{split}
\end{equation}
The non-local evolution equations for $\Upsilon_k$ and ${\cal Q}_k$ are
\begin{equation} \label{e:nonlocala}
\dot{\Upsilon}_k = \frac{1}{3}(3 \g -4 ) \Theta \Upsilon_k - \frac{4}{3} \Theta
\frac{{\cal U}}{\rho} A_k - \frac{4}{3}
\frac{{\cal U}}{\rho} \frac{k}{a} {\cal Z}_k + \frac{k}{a}
{\cal Q}_k ,
\end{equation}
and
\begin{equation} \label{e:nonlocal2a}
\dot{{\cal Q}}_k - \frac{1}{3}(3 \g - 4) \Theta
{\cal Q}_k + \frac{1}{3} \frac{k}{a} \left[\Upsilon_k + 2 \left(1
- \frac{3 K}{k^2} \right) {\cal P}_k \right]  + \frac{4}{3}\frac{k}{a}
\frac{\mathcal{U}}{\rho} A_k
=
\frac{\kappa^4}{6} \g \rho \left\{ \Theta q_k
+ \frac{k}{a} \left[ \left(1 - \frac{3 K}{k^2} \right) \pi_k - \Delta_k \right] \right\}\;.
\end{equation}
The spatial gradient of the 3-curvature scalar is
\begin{equation}
\left(\frac{k}{a} \right)^2 \left(1 - \frac{3 K}{k^2} \right) \eta_k 
= \frac{\kappa^2 \rho}{2} \Delta_k + \frac{\tilde{\kappa}^4
\rho^2}{12} \Delta_k + \frac{1}{2} \left(\frac{\tilde{\kappa}}{\kappa}
\right)^4 \rho \Upsilon_k - \frac{1}{3}\frac{k}{a} \Theta {\cal Z}_k,
\end{equation} 
and it evolves according to
\begin{equation}
\frac{k}{a} \left(1 - \frac{3 K}{k^2} \right) \left(\dot{\eta}_k - \frac{1}{3}
\Theta A_k \right) + \frac{K}{a^2} {\cal Z}_k - \frac{1}{2} \kappa^2 \rho
q_k = \frac{1}{12} \left(\frac{\tilde{\kappa}}{\kappa}\right)^4(\kappa^4
\rho^2 q_k + 6 \rho {\cal Q}_k).
\end{equation}

The evolution equations for the scalar harmonic components of the
comoving, fractional density gradients for photons, neutrinos, baryons and
cold dark matter (CDM) are
\begin{align}
\label{e:photons0}
\dot{\Delta}_{k}^{(\g)} &= -\frac{k}{a} \left(\frac{4}{3} {\cal Z}_{k} -
q_{k}^{(\g)} \right)  - \frac{4}{3} \Theta A_k \quad \text{(photons)}\;, \\
\label{e:neutrinos0}
\dot{\Delta}_{k}^{(\nu)} &= -\frac{k}{a} \left(\frac{4}{3} {\cal Z}_{k} -
q_{k}^{(\nu)} \right) - \frac{4}{3} \Theta A_k \quad \text{(neutrinos)}\;, \\
\dot{\Delta}_{k}^{(b)} &= \left(1+ \frac{P^{(b)}}{\rho^{(b)}} \right)
\label{e:baryons0}
\left[-\frac{k}{a}({\cal Z}_k - v_k^{(b)}) - \Theta A_k  \right] +
\left(\frac{P^{(b)}}{\rho^{(b)}} -  c_s^2 \right) \Theta \Delta_{k}^{(b)}
\quad \text{(baryons)}\;, \\
\label{e:CDM0}
\dot{\Delta}_{k}^{(c)} &= -\frac{k}{a} ({\cal Z}_k - v^{(c)}_{k}) -
\Theta A_k \quad \text{(CDM)}\;.
\end{align}
The evolution equations for the momentum densities and peculiar velocities
are
\begin{align}
\dot{q}^{(\g)}_k & = -\frac{1}{3}\frac{k}{a}\left[\Delta^{(\g)}_k+4A_k
+2\left(1-\frac{3K}{k^2}\right)\pi^{(\g)}_k\right] + n_e \sigma_T \left(\frac{4}{3}
v^{(b)}_k - q^{(\g)}_k \right) \; ,\\
\dot{q}^{(\nu)}_k & = -\frac{1}{3}\frac{k}{a}\left[\Delta^{(\nu)}_k+4A_k
+2\left(1-\frac{3K}{k^2}\right)\pi^{(\nu)}_k\right] \; ,\\
(\rho^{(b)}+P^{(b)}) \dot{v}^{(b)}_k &= -
(\rho^{(b)}+P^{(b)})\left[\frac{1}{3}(1-3c_s^2)\Theta v^{(b)}_k + \frac{k}{a}
A_k \right] - \frac{k}{a}(1+c_s^2) \Delta^{(b)}_k - n_e \sigma_T
\frac{\rho^{(\gamma)}}{\rho^{(\nu)}} \left(
\frac{4}{3}v^{(b)}_k - q^{(\g)}_k \right) \; , \\
\dot{v}^{(c)}_k & = - \frac{1}{3}\Theta v^{(c)}_k - \frac{k}{a}A_k \;,
\label{eq:cdmvel}
\end{align}
where the Thomson scattering terms involving the electron density
$n_e$ and Thomson cross section $\sigma_T$ arise from the interaction between
photons and the tightly-coupled baryon/electron fluid. The remaining equations
are the propagation equations for the anisotropic stresses of photons and
neutrinos, and the higher moments of their distribution functions. These
equations can be found in~\cite{challinor2}, and with polarization included
in~\cite{challinor3}, since they are unchanged from general relativity.
However, we shall not require these additional equations
at the level of approximation we make in our subsequent calculations.

\section{Perturbation dynamics in the CDM frame}

In this section we specialize our equations to FRW
backgrounds that are spatially flat\footnote{More generally,
curvature effects can be ignored for modes with wavelength much shorter
than the curvature scale, $k \gg \sqrt{|K|}$, provided the curvature does not
dominate the background dynamics.} and we
ignore the effects of the cosmological constant in the early
radiation-dominated universe. To solve the equations it is essential
to make a choice of frame $u^a$. In Ref.~\cite{challinor2} two of the
present authors adopted a frame
comoving with the CDM. Since the CDM is pressure free, this $u^a$ is geodesic
($A_a=0$) which simplifies the equations considerably. We shall adopt this
frame choice here also, though we note it may be preferable to use a frame more
closely tied to the dominant matter component over the epoch of interest. This
can be easily accomplished by adopting the energy frame ($q_a=0$). For
completeness, we give equations in the energy frame in the appendix.

We neglect baryon pressure ($c_s^2 \rightarrow 0$ and $P^{(b)} \rightarrow 0$)
and work to lowest order in the tight-coupling approximation ($n_e \sigma_T
\rightarrow \infty$; see e.g.\ Ref.~\cite{ma}). At this order the energy frame
of the photons coincides with the rest frame of the baryons, so that
$v^{(b)}_a = 3 q^{(\g)}_a /(4 \rho^{(\g)})$, and all moments of the
photon distribution are vanishingly small beyond the dipole.

With these approximations and frame choice we obtain the following equations
for the density perturbations of each component:
\begin{align}
\label{e:photons1}
\dot{\Delta}_{k}^{(\g)} &= -\frac{k}{a} \left(\frac{4}{3} {\cal Z}_{k} -
q_{k}^{(\g)} \right) \quad \text{(photons)}\;, \\
\label{e:neutrinos1}
\dot{\Delta}_{k}^{(\nu)} &= -\frac{k}{a} \left(\frac{4}{3} {\cal Z}_{k} -
q_{k}^{(\nu)} \right) \quad \text{(neutrinos)}\;, \\
\label{e:baryons1}
\dot{\Delta}_{k}^{(b)} &= -\frac{k}{a} ({\cal Z}_k - v_k^{(b)})
 \quad \text{(baryons)}\;, \\
\label{e:CDM1}
\dot{\Delta}_{k}^{(c)} &= -\frac{k}{a} {\cal Z}_k \quad \text{(CDM)}\;.
\end{align}
The equations for the peculiar velocities and momentum densities are
\begin{align}
(4\rho^{(\g)} + 3 \rho^{(b)}) \dot{q}^{(\g)}_k & =
-\frac{4}{3}\frac{k}{a} \rho^{(\g)}\Delta^{(\g)}_k - \rho^{(b)}\Theta
q^{(\g)}_k \; , \label{e:tightcouple1}\\
\dot{q}^{(\nu)}_k & = -\frac{1}{3}\frac{k}{a}\left(\Delta^{(\nu)}_k
+2\pi^{(\nu)}_k\right) \; ,
\end{align}
along with $v^{(c)}_k = 0$ and $v^{(b)}_k = 3 q^{(\g)}_k /4$. The latter
equation, together with Eqs.~(\ref{e:photons1}) and (\ref{e:baryons1}),
implies that $\dot{\Delta}^{(b)}_k = 3 \dot{\Delta}^{(\g)}_k /4$ so
that any entropy perturbation between the photons and baryons is conserved
while tight coupling holds.
The effects of baryon inertia appear in Eq.~(\ref{e:tightcouple1})
because of the tight coupling between the baryons and photons.

The constraint equations are found to be:
\begin{equation} \label{e:constraint1c}
\kappa^2 \rho q_k - \frac{2}{3} \left(\frac{k}{a} \right)^2 ({\cal Z}_k
  - \sigma_k ) - \frac{1}{6}
  \left(\frac{\tilde{\kappa}}{\kappa}\right)^4 (\kappa^4 \rho^2 q_k + 6 \rho
  {\cal Q}_k) = 0\;,
\end{equation}
and
\begin{equation} \label{e:constraint2c}
\begin{split}
& 2 \left(\frac{k}{a} \right)^3 \Phi_k -
\kappa^2 \rho \left(\frac{k}{a} \right) (\Delta_k - \pi_k ) + \kappa^2
\Theta \rho q_k \\
&= \frac{1}{16}\left(\frac{\tilde{\kappa}}{\kappa}\right)^4
\bigg[\kappa^4 \left(-\frac{8}{3} \Theta \rho^2 q_k + \frac{4}{3} \frac{k}{a}
\rho^2 \left[(3\g -2)
 \pi_k  +  2 \Delta_k \right] \right)  \\
& + 16  \left(\frac{k}{a} \right) \rho (\Upsilon_k -{\cal P}_k)  - 16
\Theta \rho {\cal Q}_k \bigg]\;.
\end{split}
\end{equation}
The propagation equation for the comoving expansion gradient in the
CDM frame is
\begin{equation} \label{e:propagation3c}
\begin{split}
&\dot{{\cal Z}}_k + \frac{1}{3} \Theta \mathcal{Z}_k + \frac{\kappa^2}{2}
\frac{a}{k} \left[2 (\rho^{(\g)}
\Delta^{(\g)}_k + \rho^{(\nu)} \Delta^{(\nu)}_k ) + \rho^{(b)}
\Delta^{(b)}_k + \rho^{(c)} \Delta^{(c)}_k  \right] \\
&= -\frac{1}{12}  \left(\frac{\tilde{\kappa}}{\kappa}\right)^4 \frac{a}{k}
\left\{\kappa^4 [(2\rho+3P)\rho\Delta_k + \rho(3\rho^{(\g)}\Delta^{(\g)}_k
+3\rho^{(\nu)}\Delta^{(\nu)}_k + (2+3c_s^2)\rho^{(b)}\Delta^{(b)}_k
+ 2 \rho^{(c)}\Delta^{(c)}_k)]+12 \rho \Upsilon_k \right\}\;.
\end{split}
\end{equation}
The variables $\Phi_k$ and $\sigma_k$ can be determined from the constraint
equations so their propagation equations are not independent of the above
set. The propagation equation for $\Phi_k$ is unchanged from
Eq.~$\eqref{e:propagation1a}$ since that equation was already written in
frame-invariant form. The propagation equation for the shear in the CDM frame
is
\begin{equation}
\frac{k}{a} \left(\dot{\sigma}_k + \frac{1}{3} \Theta \sigma_k \right) +
\left(\frac{k}{a}\right)^2 \Phi_k - \frac{\kappa^2}{2} \rho \pi_k =
\frac{1}{24}\left(\frac{\tilde{\kappa}}{\kappa}\right)^4 [-(3 \g - 2) \kappa^4
\rho^2 \pi_k + 12 \rho {\cal P}_k]\;.
\end{equation}

Finally we have the non-local evolution equations for ${\Upsilon}_k$ and
${\cal Q}_k$ which in the CDM frame become
\begin{equation} \label{e:nonlocal1c}
\dot{\Upsilon}_k = \frac{1}{3}(3 \g -4 )\Theta
\Upsilon_k - \frac{4}{3} \frac{{\cal U}}{\rho} \frac{k}{a}
{\cal Z}_k + \frac{k}{a} {\cal Q}_k \;,
\end{equation}
and
\begin{equation} \label{e:nonlocal2c}
\dot{{\cal Q}}_k -\frac{1}{3} (3 \g - 4) \Theta {\cal Q}_k + \frac{1}{3}
\frac{k}{a} (\Upsilon_k +2 \mathcal{P}_k)=
\frac{\kappa^4}{6} \g \rho \left[ \Theta q_k + \frac{k}{a}
(\pi_k - \Delta_k) \right]\;.
\end{equation}

\section{Solutions in the radiation-dominated era} \label{sec:radiation}

We now use the above equations to extract the
mode solutions of the scalar perturbation equations in the
radiation-dominated era, $\g=4/3$. To simplify matters, as well
as neglecting the contribution of the baryons and CDM to the background
dynamics, we shall only consider those modes for which $D_a \rho^{(b)}$
and $D_a \rho^{(c)}$ make a negligible contribution to the total
matter perturbation $D_a \rho$. This approximation allows us to write
the total matter perturbations in the form
\begin{equation}
(\rho^{(\g)} + \rho^{(\nu)})\Delta_k = \rho^{(\g)}\Delta^{(\g)}_k
+ \rho^{(\nu)} \Delta^{(\nu)}_k, \quad
(\rho^{(\g)} + \rho^{(\nu)}) q_k = \rho^{(\g)}q^{(\g)}_k
+ \rho^{(\nu)} q^{(\nu)}_k,
\label{eq:approx}
\end{equation}
and effectively removes the back-reaction of the baryon and CDM perturbations
on the perturbations of the spacetime geometry. We note that in making this
approximation we lose two modes corresponding to the baryon and CDM
isocurvature (density) modes of general relativity, in which the sub-dominant
matter components make significant contributions to the total fractional
density perturbation (which vanishes as $t\rightarrow 0$). However, for
our purposes the loss of generality is not that important, while the
simplifications resulting from decoupling the baryon and photon perturbations
are considerable. We also neglect moments of the neutrino distribution
function above the dipole (so there is no matter anisotropic stress). This
approximation is good for super-Hubble modes, but fails due to neutrino
free streaming on sub-Hubble scales.

We shall also assume that the non-local
energy density ${\cal U}$ vanishes in the background for all
energy regimes \cite{gordon}. Physically, vanishing ${\cal U}$
corresponds to the background bulk being conformally flat and strictly
Anti-de Sitter. Note that ${\cal U}=0$ in the background need not imply that 
the fluctuations in the non-local
energy density are zero, i.e.\ $\Upsilon_a \neq 0$.

With the above conditions the
following set of equations are obtained: 
\begin{align}
\label{e:ae1}
\left(\frac{k}{a} \right)^2 (\dot{\Phi}_k + H \Phi_k ) + \frac{\kappa^2
\rho}{2} \left(\frac{k}{a} \right) \left( \frac{4}{3} \sigma_k - q_k \right) \left( 1 +
\frac{\rho}{\lambda} \right) &= \frac{3}{\kappa^2} \frac{\rho}{\lambda}
\left[ \left(\frac{k}{a} \right) {\cal Q}_k + 3 H {\cal P}_k -
\dot{{\cal P}}_k
\right]\;, \\
\label{e:ae2}
\left(\frac{k}{a} \right) (\dot{{\cal Z}}_k + H
{\cal Z}_k) + \kappa^2 \rho \left(1+ \frac{3 \rho}{\lambda} \right) \Delta_k &= -
\frac{6}{\kappa^2} \frac{\rho}{\lambda} \Upsilon_k \\
\label{e:ae3}
\left(\frac{k}{a} \right) (\dot{\sigma}_k + H \sigma_k) + \left(\frac{k}{a}
\right)^2 \Phi_k &= \frac{3}{\kappa^2} \frac{\rho}{\lambda}
{\cal P}_k\;,  \\
\label{e:ae4}
\dot{q}_k^{(\g)} + \frac{1}{3} \frac{k}{a} \Delta_k^{(\g)} &=0\;, \\
\label{e:ae5}
\dot{q}_k^{(\nu)} + \frac{1}{3} \frac{k}{a} \Delta_k^{(\nu)} &=0\;, \\
\label{e:ae6}
\dot{\Delta}_{k}^{(\g)} + \frac{k}{a} \left(\frac{4}{3} {\cal Z}_{k} -
q_{k}^{(\g)} \right) &=0\;, \\
\label{e:ae7}
\dot{\Delta}_{k}^{(\nu)} +\frac{k}{a} \left(\frac{4}{3} {\cal Z}_{k} -
q_{k}^{(\nu)} \right) &=0,
\end{align}
where recall $H=\Theta/3$. For the constraint equations we find
\begin{align}
\label{e:ae8}
3 \kappa^2 \left(1 + \frac{\rho}{\lambda} \right) \rho q_k - 2
\left(\frac{k}{a}\right)^2 ({\cal Z}_k - \sigma_k) &= -\frac{18}{\kappa^2}
\frac{\rho}{\lambda} {\cal Q}_k\;, \\
\label{e:ae9}
2 \left(\frac{k}{a}\right)^3 \Phi_k + \kappa^2 \rho \left(1 + \frac{\rho}{\lambda}
\right) \left[ 3 H q_k - \left(\frac{k}{a} \right) \Delta_k \right] &= \frac{6}{\kappa^2}
\frac{\rho}{\lambda} \left[\left( \frac{k}{a} \right) (\Upsilon_k -
{\cal P}_k) - 3H {\cal Q}_k \right]\;.
\end{align}
Finally the non-local evolution equations are found to be :
\begin{align}
\label{e:ae10}
\dot{\Upsilon}_k &=   \frac{k}{a} {\cal Q}_k\;, \\
\label{e:ae11}
9 \dot{{\cal Q}}_k + 3 \left( \frac{k}{a} \right) (\Upsilon_k + 2
{\cal P}_k) &= -2 \kappa^4 \rho \left(\frac{k}{a} \Delta_k - 3H
q_k \right)\;.
\end{align}
It is easy to show by propagating the constraint equations
that the above set of equations are consistent. 

By inspection, there is a solution of these equations with
\begin{align}
\Phi_k &= 0 , \\
{\cal Z}_k &= \left[3 \dot{H}
\left(\frac{a}{k}\right)^2 -1\right]\frac{A}{a} ,\\
\sigma_k &= - \frac{A}{a} , \\
q^{(\gamma)}_k &= - \frac{4}{3} \frac{A}{a} ,\\
q^{(\nu)}_k &= - \frac{4}{3} \frac{A}{a} ,\\
\Delta^{(\gamma)}_k &= -4 H \frac{A}{k} , \\
\Delta^{(\nu)}_k &= -4 H \frac{A}{k} , \\
\Upsilon_k &= 0 , \\
{\cal Q}_k &= 0 , \\
{\cal P}_k &=0,
\end{align} 
where $A$ is a constant.
This solution describes a radiation-dominated universe that is exactly FRW
except that the CDM has a peculiar velocity $\bar{v}_a^{(c)} = (A/a)
Q_a^{(k)}$ relative to the velocity of the FRW
fundamental observers. [This form for $\bar{v}_a^{(c)}$ clearly satisfies
Eq.~(\ref{eq:cdmvel}) with $A_a=0$.] Such a solution is possible since we have
neglected the gravitational effect of the CDM (and baryon) perturbations
in making the approximations in Eq.~(\ref{eq:approx}). The same solution arises
in general relativity~\cite{challinor2}. Including the back-reaction of the
CDM perturbations, we would find additional small peculiar velocities in the
dominant matter components which compensate the CDM flux. We shall not
consider this irregular CDM isocurvature velocity mode any further here.

Another pair of solutions are easily found by decoupling the photon/neutrino
entropy perturbations. Introducing the photon/neutrino entropy
perturbation (up to constant) $\Delta_2$ and relative flux $q_2$:
\begin{equation}
\begin{split}
\Delta_2 &=  \Delta^{(\g)}_k - \Delta^{(\nu)}_k\;, \\
q_2 &= q^{(\g)}_k - q^{(\nu)}_k\;,
\end{split}
\end{equation}
the equations for $\Delta_2$ and $q_2$ decouple to give
\begin{align}
\dot{\Delta}_2 - \frac{k}{a} q_2 &= 0 \;,\\
\dot{q}_2 + \frac{1}{3} \frac{k}{a} \Delta_2 &=0 \;.
\end{align}
Switching to conformal time ($d \tau = dt /a$) we can solve for $\Delta_2$ and
$q_2$ to find
\begin{align}
q_2 (\tau) &= B \cos \left(\frac{k \tau}{3} \right) + C \sin \left(\frac{k
\tau}{3} \right)\;, \\ 
\Delta_2 (\tau) &= B \sin \left(\frac{k \tau}{3} \right)  - C \cos \left(\frac{k
\tau}{3} \right) \;.
\end{align}
The constants $B$ and $C$ label the neutrino velocity and density isocurvature
modes respectively~\cite{challinor2,bucher}, in which the neutrinos and photons
initially have mutually compensating peculiar velocities and density
perturbations. The perfect decoupling of these
isocurvature modes is a consequence of our neglecting anisotropic stresses
(and higher moments of the distribution functions) and baryon inertia.

Having decoupled the entropy perturbations, we write the remaining equations
in terms of the total variables $\Delta_k$ and $q_k$. The propagation equations
for the non-local variables $\Upsilon_k$ and ${\cal Q}_k$ are redundant since
these variables are determined by the constraint equations~(\ref{e:ae8}) and
(\ref{e:ae9}):
\begin{align} \label{eq:cons1}
\frac{6}{\kappa^2} \frac{\rho}{\lambda} \Upsilon_k &= 2
\left(\frac{k}{a}\right)^2 \Phi_k + 2H \left(\frac{k}{a} \right) ({\cal Z}_k -
\sigma_k) - \kappa^2 \rho \left( 1+ \frac{\rho}{\lambda} \right) \Delta_k 
+ \frac{6}{\kappa^2} \frac{\rho}{\lambda} {\cal P}_k, \\
\frac{3}{\kappa^2} \frac{\rho}{\lambda} {\cal Q}_k &= \frac{1}{3}
\left(\frac{k}{a} \right)^2 ({\cal Z}_k - \sigma_k) - \frac{\kappa^2 \rho}{2}
\left( 1+ \frac{\rho}{\lambda} \right) q_k. \label{eq:cons2} 
\end{align}
Substituting these expressions in the right-hand sides of Eqs.~(\ref{e:ae1})
and (\ref{e:ae2}) we find 
\begin{align}
\label{e:be1}
\left(\frac{k}{a}\right)^2 \left(\dot{\Phi}_k + H \Phi_k \right) + \frac{2\kappa^2
\rho}{3} \left(\frac{k}{a} \right) \left(1+\frac{\rho}{\lambda}\right)
\sigma_k - \frac{1}{3} \left(\frac{k}{a}\right)^3
({\cal Z}_k - \sigma_k) &= \frac{3}{\kappa^2} \frac{\rho}{\lambda} (3H
{\cal P}_k - \dot{\cal P}_k) \;,\\
\label{e:be2}
\left(\frac{k}{a} \right) \dot{{\cal Z}}_k + H\left(\frac{k}{a} \right)
{\cal Z}_k + \kappa^2 \rho \left(\frac{2 \rho}{\lambda} \right) \Delta_k +
2\left(\frac{k}{a}\right)^2 \Phi_k + 2H \left(\frac{k}{a}\right)({\cal Z}_k -
\sigma_k)  &= - \frac{6 \rho}{\kappa^2 \lambda}
{\cal P}_k \;,\\
\label{e:be3}
\left(\frac{k}{a} \right) (\dot{\sigma}_k + H \sigma_k) + \left(\frac{k}{a}
\right)^2 \Phi_k &= \frac{3}{\kappa^2} \frac{\rho}{\lambda}
{\cal P}_k\;,  \\
\label{e:be4}
\dot{\Delta}_k + \frac{k}{a} \left(\frac{4}{3} {\cal Z}_k - q_k \right) &=0 \;,
\\
\label{e:be5}
\dot{q}_k + \frac{1}{3} \frac{k}{a} \Delta_k &=0 \;.
\end{align}
These equations describe the evolution of the intrinsic perturbations
to the brane. The usual general relativistic constraint equations are now
replaced by the constraints~(\ref{eq:cons1}) and (\ref{eq:cons2}) which
determine two of the non-local variables. The lack of a propagation equation
for ${\cal P}_k$ reflects the incompleteness of the 1+3 dimensional
description of braneworld dynamics.

In the following it will prove convenient to adopt the dimensionless
independent variable
\begin{equation} \label{e:trans1}
x= \frac{k}{Ha}\;,
\end{equation}
which is (to within a factor of $2\pi$) the ratio of the Hubble length
to the wavelength of the perturbations. Using the (modified) Friedmann
equations for the background in radiation domination, and with
$\mathcal{U}=0$, we find that
\begin{equation}
\frac{dx}{dt} = \frac{k}{a}\left(\frac{2+3\rho/\lambda}{2+\rho/\lambda}\right).
\end{equation}
The relative importance of the local (quadratic) braneworld corrections
to the Einstein equation depends on the dimensionless ratio $\rho/\lambda$.
In the low-energy limit, $\rho\ll \lambda$, the quadratic local corrections can
be neglected although the non-local corrections $\mathcal{E}_{ab}$ may
still be important. In the opposite (high-energy) limit the quadratic
corrections dominate over the terms that are linear in the energy-momentum
tensor. We now consider these two limits separately.

\subsection{Low-energy regime}
\label{sec:lowenergy}

In the low-energy regime we have $dx/dt \approx k/a$ and $x \approx k \tau$.
The total energy density $\rho$ is proportional to $x^{-4}$. Denoting
derivatives with respect to $x$ with a prime, using $\rho \ll \lambda$,
and assuming that we can neglect the term involving $(\rho/\lambda)\Delta_k$
in Eq.~(\ref{e:be2}) compared to the other terms, we find
\begin{align}
\label{e:le1}
3x^2 \Phi_k'+ 3x \Phi_k + (6+x^2) \sigma_k - x^2 {\cal Z}_k &=
\frac{27}{\kappa^4 \lambda} (3 {\cal P}_k - x {\cal P}_k') \\
\label{e:le2}
x^2 {\cal Z}_k' + 3 x {\cal Z}_k + 2x^2 \Phi_k - 2x \sigma_k &=
- \frac{18}{\kappa^4 \lambda} {\cal P}_k \\
\label{e:le3}
x^2 \sigma_k' + x\sigma_k + x^2 \Phi_k &=
\frac{9}{\kappa^4 \lambda} {\cal P}_k \\
\label{e:le4}
\Delta_k' + \frac{4}{3} {\cal Z}_k - q_k &=0 \\
\label{e:le5}
q_k' + \frac{1}{3} \Delta_k &=0 
\end{align}
Combining these equations we find an inhomogeneous, second-order equation
for $\Phi_{k}$:
\begin{equation}
3x \Phi_k'' + 12 \Phi_k' + x \Phi_k = F_k(x),
\label{eq:secondPhi}
\end{equation}
where
\begin{equation}
F_k(x) \equiv -\frac{27}{\kappa^4 \lambda}
\left[ {\cal P}_k'' - \frac{{\cal P}_k'}
{x} + \left(\frac{2}{x^3} - \frac{3}{x^2} + \frac{1}{x} \right) {\cal P}_k \right].
\end{equation}
In general relativity the same second-order equation holds for $\Phi_k$ but
with $F_k(x)=0$.

The presence of terms involving the non-local anisotropic stress on the
right-hand side of Eq.~(\ref{eq:secondPhi}) ensure that $\Phi_k$ cannot be
evolved on the brane alone. The resolution of this problem will require
careful analysis of the bulk dynamics in five dimensions. In this paper
our aims are less ambitious; we shall solve Eq.~(\ref{eq:secondPhi}) with
${\cal P}_k=0$.
Although we certainly do not expect ${\cal P}_{ab}=0$\footnote{We
have not investigated the consistency of the condition ${\cal P}_{ab}=0$
with the five-dimensional bulk dynamics in the presence of a perturbed brane.},
the solutions of the homogeneous equation may still prove a useful starting
point for a more complete analysis. For example, they allow one to construct
Green's functions for Eq.~(\ref{eq:secondPhi}) which could be used to assess
the impact of specific ansatze for ${\cal P}_{ab}$~\cite{barrow}.

With ${\cal P}_k=0$ we can solve Eqs.~(\ref{e:le1})--(\ref{e:le5})
analytically to find
\begin{align}
\Phi_k &= \frac{c_1}{x^3} \left[3 \sin \left(\frac{x}{\sqrt{3}} \right) -
x\sqrt{3} \cos \left(\frac{x}{\sqrt{3}} \right) \right] 
+ \frac{c_2}{x^3} \left[3 \cos  \left(\frac{x}{\sqrt{3}} \right)+ x \sqrt{3}
\sin \left(\frac{x}{\sqrt{3}} \right) \right] , \\
\sigma_k &= \frac{3}{x^2} \left[c_2 \cos  \left(\frac{x}{\sqrt{3}} \right)
+ c_1 \sin  \left(\frac{x}{\sqrt{3}} \right) \right] + \frac{c_3}{x}, \\
{\cal Z}_k &= \frac{c_3 (6+x^2)}{x^3} + \frac{6 \sqrt{3}}{x^3} \left[c_1
\cos\left(\frac{x}{\sqrt{3}} \right)  - c_2 \sin \left(\frac{x}{\sqrt{3}}
\right)\right] +\frac{6}{x^2} \left[c_2 \cos \left(\frac{x}{\sqrt{3}} \right)
+ c_1 \sin \left(\frac{x}{\sqrt{3}} \right) \right], \\
\Delta_k &= c_4 \cos\left(\frac{x}{\sqrt{3}}\right)  + c_5 \sin
\left(\frac{x}{\sqrt{3}}\right)
+ \frac{4 c_3}{x^2} + \frac{4}{x}
\left[c_2 \cos\left(\frac{x}{\sqrt{3}} \right)  + c_1 \sin \left(\frac{x}
{\sqrt{3}} \right) \right] \notag \\ 
& \mbox{} + \left(\frac{4 \sqrt{3}}{x^2}-\frac{2}{\sqrt{3}}\right)
\left[c_1 \cos \left(\frac{x}{\sqrt{3}} \right)
- c_2 \sin\left(\frac{x}{\sqrt{3}} \right)  \right] , \\
q_k &= \frac{c_5}{\sqrt{3}} \cos \left(\frac{x}{\sqrt{3}} \right) -
\frac{c_4}{\sqrt{3}} \sin\left(\frac{x}{\sqrt{3}} \right)    + \frac{4
c_3}{3x} + \frac{4x}{\sqrt{3}} \left[c_1 \cos \left(\frac{x}{\sqrt{3}}\right)
- c_2 \sin \left(\frac{x}{\sqrt{3}} \right) \right] \notag \\
& \mbox{} + \frac{2}{3}\left[c_2 \cos \left(
\frac{x}{\sqrt{3}}\right) + c_1 \sin \left(\frac{x}{\sqrt{3}}\right) \right].
\end{align}
The mode labelled by $c_3$ is the CDM velocity isocurvature mode discussed
earlier. The modes labelled by $c_1$ and $c_2$ are the same as in general
relativity; they describe the adiabatic growing and decaying solutions
respectively. However, in the low-energy limit we also find two additional
isocurvature modes ($c_4$ and $c_5$) that are not present in general
relativity. These
arise from the two additional degrees of freedom $\Upsilon_k$ and ${\cal Q}_k$
present in the braneworld model (with $P_k=0$). The mode $c_4$ initially
has non-zero but compensating gradients in the total matter and
non-local densities, and $c_5$ initially has compensated energy fluxes.
Formally these isocurvature solutions violate the assumption that the
term involving $(\rho/\lambda)\Delta_k$ be negligible compared to the
other terms in Eq.~(\ref{e:be2}) since all other terms vanish. In practice,
there will be some gravitational back-reaction onto the other gauge-invariant
variables controlled by the dimensionless quantity $\rho/\lambda$, but the
general character of these isocurvature modes will be preserved for
$\rho/\lambda\ll 1$.

\subsection{High-energy regime}

We now turn to the high-energy regime, where the quadratic terms in
the stress-energy tensor dominate the (local) linear terms. In this limit
the scale factor $a \propto t^{1/4}$. The modification to the expansion rate
leads to an increase in the amplitude of scalar and tensor fluctuations 
produced during high-energy inflation~\cite{barrow}. With
${\cal U}=0$ in the background, and $\rho \gg \lambda$, the Hubble parameter
is approximately
\begin{equation}
H^2 \approx \frac{1}{36} \tilde{\kappa}^4 \rho^2,
\end{equation}
and $dx/dt \approx 3 k/a$. In terms of conformal time $\tau$, $x \approx
3 k \tau$.
  
\subsubsection{Power series solutions for the high-energy regime}
\label{power}

It is convenient to rescale the non-local variables by the dimensionless
quantity $\kappa^4 \rho$. Thus we define
\begin{align}
\label{e:trans2}
\bar{\Upsilon}_k &\equiv \frac{\Upsilon_k}{\kappa^4 \rho}\;, \\
\label{e:trans3}
\bar{{\cal Q}}_k &\equiv \frac{{\cal Q}_k}{\kappa^4 \rho}\;, \\
\label{e:trans4}
\bar{{\cal P}}_k &\equiv \frac{{\cal P}_k}{\kappa^4 \rho}\;.
\end{align}
The fractional total (effective) density perturbation and energy flux
can be written in terms of the barred variables
[e.g.\ $\bar{\Upsilon}_a \equiv \Upsilon_a/ (\kappa^4 \rho)$] in the
high-energy limit as
\begin{eqnarray}
\frac{a D_a \rho^{\text{tot}}}{\rho^{\text{tot}}} &\approx& 2(\Delta_a + 6
\bar{\Upsilon}_a), \\
q_a^{\text{tot}} &\approx& \frac{2\rho^{\text{tot}}}{\rho} (q_a + 6
\bar{\cal Q}_a ).
\end{eqnarray}
Making the high-energy approximation $\rho \gg \lambda$ in
Eqs.~$\eqref{e:be1}$--$\eqref{e:be5}$, we obtain  
\begin{eqnarray}
\label{e:he2a}
9x^2 \Phi^{'}_k + 3x \Phi_k + (12+x^2) \sigma_k - x^2 {\cal Z}_k  &=&
54 \left[ \frac{7 \bar{\cal P}_k}{x} - 3 \bar{\cal P}_k' \right]\;,  \\
\label{e:he2b}
3x^2 {\cal Z}^{'}_k + 3x {\cal Z}_k - 2x \sigma_k + 2x^2
\Phi_k + 12 \Delta_k &=& -36 \bar{\cal P}_k\;, \\
\label{e:he2c}
3x \sigma^{'}_k + \sigma_k + x \Phi_k &=& 18 \frac{\bar{{\cal P}}_k}{x}\;,
\\
\label{e:he2d}
\Delta_k^{'} - \frac{1}{3} q_k  + \frac{4}{9} {\cal Z}_k &=&0\;, \\
\label{e:he2e}
q_k^{'} + \frac{1}{9} \Delta_k &=&0\;.
\end{eqnarray}
The non-local quantities $\bar{\Upsilon}_k$ and $\bar{{\cal Q}}_k$ are
determined by the constraints
\begin{eqnarray}
\bar{\Upsilon}_k &=& \frac{1}{18}x^2 \Phi_k + \frac{1}{18} x
({\cal Z}_k-\sigma_k) - \frac{1}{6} \Delta_k + \bar{{\cal P}}_k, \\
\bar{{\cal Q}}_k &=& \frac{1}{54} x^2 ({\cal Z}_k - \sigma_k) - \frac{1}{6}
q_k.
\end{eqnarray}

We can manipulate Eqs.~$\eqref{e:he2a}$--$\eqref{e:he2e}$ to obtain
a fourth-order equation for the gravitational potential $\Phi_k$:
\begin{equation} \label{e:order4de}
729 x^2 \frac{\p^4 \Phi_k}{\p x^4} +3888 x \frac{\p^3 \Phi_k}{\p x^3} +
(1782+54x^2) \frac{\p^2\Phi_k}{\p x^2}
+144x \frac{\p\Phi_k}{\p x} + (90+x^2) \Phi_k = F_k(x)\;,
\end{equation}
where
\begin{equation} \label{e:p-de1}
F_k(x) = -\frac{54}{x^4}\left( 243 x^4 \frac{\partial^4 \bar{\cal P}_k}{
\partial x^4} - 810 x^3 \frac{\partial^3 \bar{\cal P}_k}{\partial x^3}
+18x^2(135+2x^2) \frac{\partial^2 \bar{\cal P}_k}{\partial x^2}
-30 x(162+x^2) \frac{\partial \bar{\cal P}_k}{\partial x}
+ [x^4 + 30(162+x^2)]\bar{\cal P}_k\right)\; .
\end{equation}
Since we do not have an evolution equation for $\bar{\cal P}_k$ we
adopt the strategy taken in the low-energy limit and look for solutions
of the homogeneous equations ($\bar{{\cal P}}_k=0$).
In principle one can use these solutions to construct formal solutions
of the inhomogeneous equations with Green's method.

To solve Eq.~(\ref{e:order4de}) with $\bar{{\cal P}}_k=0$ we
construct a power series solution for $\Phi_k(x)$:
\begin{equation}
\Phi_k(x) = x^m \sum^{\infty}_{n=0} a_n x^n\;,
\end{equation}
where $a_0 \neq 0$. The indicial equation for $m$ is
\begin{equation} \label{e:indicial}
m (m-1) (3m+5) (3m-4)=0\;.
\end{equation}
For each value of $m$ we substitute into Eq.~(\ref{e:order4de}) and solve
the resulting recursion relations for the $\{a_n\}$. We then obtain the
other gauge-invariant variables by direct integration. The original set of
equations~(\ref{e:he2a})--(\ref{e:he2e}) has five degrees of freedom, so we
expect one additional solution with $\Phi_k=0$. This solution is the CDM
isocurvature solution discussed earlier, and has a finite series expansion:
\begin{eqnarray}
\Phi_k &=& 0\;, \\
\sigma_k  &=& C x^{-\frac{1}{3}}\;, \\
{\cal Z}_k &=& C x^{-\frac{7}{3}} (12 + x^2)\;, \\
\Delta_k &=& 4 C x^{-\frac{4}{3}}\;, \\
q_k &=& \frac{4}{3} C x^{-\frac{1}{3}}\;,
\end{eqnarray}
where $C$ is a constant. The non-local variables vanish.

The first two terms of the mode with $m=0$ are
\begin{eqnarray}
\Phi_k &=& b_1 \left(1 - \frac{5}{198} x^2 \right)\;, \\
\sigma_k &=& b_1 \left(-\frac{1}{4} x + \frac{1}{396} x^3 \right)\;,    \\
{\cal Z}_k &=& b_1 \left(- \frac{3}{4} x + \frac{5}{864} x^3 \right)\;, \\
\Delta_k &=& b_1 \left(\frac{1}{6} x^2 - \frac{1}{864} x^4 \right)\;,  \\
q_k &=&  b_0 \left(-\frac{1}{162} x^3 + \frac{1}{38880} x^5 \right)\;, \\
\bar{\Upsilon}_k &=& b_1 \left(-\frac{1}{972} x^4
+ \frac{1}{249480} x^6 \right) \;, \\
\bar{{\cal Q}}_k &=& b_1 \left(-\frac{2}{243} x^3 + \frac{1}{17820} x^5
\right)\;,
\end{eqnarray}
where $b_1$ is a constant.
The form of this solution is similar to the adiabatic growing mode of
general relativity.

The mode corresponding to $m=1$ is
\begin{eqnarray}
\Phi_k &=& b_2 \left(x - \frac{13}{1890} x^3 \right)\;, \\
\sigma_k &=& b_2 \left(-\frac{1}{7} x^2 + \frac{1}{1890} x^4 \right)\;, \\
{\cal Z}_k &=& b_2 \left(\frac{72}{7} - \frac{12}{35} x^2 \right)\;,  \\
\Delta_k  &=& b_2 \left(-\frac{18}{7} x + \frac{1}{15} x^3 \right)\;, \\
q_k  &=& b_2  \left(6 + \frac{1}{7} x^2 \right)\;, \\
\bar{\Upsilon}_k &=& b_2 \left(x + \frac{1}{30} x^3 \right)\;, \\
\label{e:Qiso}
\bar{{\cal Q}}_k &=& b_2 \left(-1 + \frac{1}{6} x^2  \right)\;,
\end{eqnarray}
with $b_2$ a constant. As $t \rightarrow 0$ there are non-zero but
compensating contributions to the effective peculiar velocity $q_a^{\text{tot}}
/\rho^{\text{tot}}$ from the matter and the non-local energy fluxes. The
contributions of these components to the fractional total density perturbation
$a D_a \rho^{\text{tot}}/\rho^{\text{tot}}$ vanish as
$t \rightarrow 0$. It follows
that this solution describes an isocurvature velocity mode where the early
time matter and non-local (Weyl) components have equal but opposite peculiar
velocities in the CDM frame. The existence of such isocurvature modes
was predicted in Refs~\cite{langlois} and \cite{gordon} for large-scale density
perturbations.  

The mode corresponding to $m=-\frac{5}{3}$ is singular as $t \rightarrow 0$
(it is a decaying mode):
\begin{eqnarray}
\Phi_k &=& b_3 x^{-\frac{5}{3}} \left( 1 - \frac{5}{18} x^2 \right)\;, \\
\sigma_k &=& b_3 x^{-\frac{2}{3}} \left( 1 + \frac{1}{18} x^2 \right)\;, \\
{\cal Z}_k &=& b_3 \left(\frac{14}{99} x^{\frac{4}{3}} - \frac{1217}{1590435}
x^{\frac{10}{3}} \right)\;, \\
\Delta_k  &=& b_3 \left(-\frac{8}{297} x^{\frac{7}{3}} + \frac{64}{433755}
x^{\frac{13}{3}} \right)\;, \\
q_k  &=& b_3 \left(\frac{4}{4455} x^{\frac{10}{3}} - \frac{4}{1301265}
x^{\frac{16}{3}} \right)\; ,\\
\bar{\Upsilon}_k &=&  b_3 \left(-\frac{1}{162} x^{\frac{7}{3}} +
\frac{7}{43740} x^{\frac{13}{3}} \right)\;,  \\
\bar{{\cal Q}}_k &=& b_3 \left(-\frac{1}{54} x^{\frac{4}{3}} +
\frac{7}{4860} x^{\frac{10}{3}} \right)\;.
\end{eqnarray}
A similar mode is found in general relativity but there the decay of $\Phi_k$
is more rapid ($\Phi_k \propto x^{-3}$) on large scales.

Finally, for $m=\frac{4}{3}$ we have
\begin{eqnarray}
\Phi_k &=& b_4 x^{\frac{4}{3}} \left(1 - \frac{17}{3150} x^2 \right)\;, \\
\sigma_k &=& b_4 x^{\frac{4}{3}} \left(- \frac{1}{8} x + \frac{17}{44100}
x^3 \right)\;, \\
{\cal Z}_k &=& b_4 x^{\frac{1}{3}} \left(\frac{27}{2} - \frac{117}{392} x^2 \right)
\;, \\
\Delta_k  &=& b_4 x^{\frac{4}{3}} \left(-\frac{9}{2} + \frac{3}{49} x^2
\right)\;, \\
q_k &=& b_4 x^{\frac{4}{3}} \left(\frac{3}{14} x - \frac{1}{637} x^3
\right)\; ,\\
\bar{\Upsilon} (x) &=& b_4 x^{\frac{4}{3}} \left(\frac{3}{2} + \frac{1}{28}
x^2 \right)\;, \\
\bar{{\cal Q}} (x) &=& b_4 x^{\frac{4}{3}} \left(\frac{3}{14} x -
\frac{29}{9828} x^3 \right)\;.
\end{eqnarray}
In this mode the universe asymptotes to an FRW (brane) model in the past
as $t \rightarrow 0$. Note that this requires careful cancellation
between $a D_a \rho^{\text{tot}} / \rho^{\text{tot}}$ and $q_a^{\text{tot}}$
to avoid a singularity in the gravitational potential $\Phi_k$ (which
would diverge as $x^{-2/3}$ without such cancellation). Like the velocity
isocurvature mode ($m=1$) discussed above, this mode has no analogue in
general relativity.

\section{A covariant expression for the temperature anisotropy}
\label{sec:aniso}

In this section we discuss the line of sight solution to the
Boltzmann equation for the scalar contribution to the gauge-invariant
temperature anisotropy $\delta_T(e)$ of the CMB in braneworld models.
We employ the 1+3 covariant approach, and show that our result is equivalent
to that given recently by
Langlois et al~\cite{langlois} using the
Bardeen formalism.

Over the epoch of interest the individual matter
constituents of the universe interact with each other under gravity only,
except for the photons and baryons (including the electrons), which
are also coupled through Thomson scattering. The variation of
the gauge-invariant temperature perturbation $\delta_T (e)$, where $e^a$ is
the (projected) photon propagation direction, along the
line of sight is given by the (linearized) covariant Boltzmann
equation (valid for scalar, vector, and tensor modes)~\cite{challinor1}:
\begin{equation} \label{e:temperature1}
\begin{split}
\delta_T (e)' + \sigma_T n_e \delta_T (e) &= -\sigma_{ab} e^a e^b - A_a
e^a - \frac{e^a D_a \rho^{(\gamma)}}{4\rho^{(\gamma)}} -
\frac{D^a q^{(\gamma)}_a}{4\rho^{(\gamma)}} \\
& + \sigma_T n_e \left(v^{(b)}_a e^a + \frac{3}{16} \rho^{(\g)}
\pi_{ab}^{(\g)} e^a e^b \right)\;,
\end{split}
\end{equation}
where the prime denotes the derivative with respect to a parameter $\lambda$
defined along the line of sight by $d \lambda = - u_a dx^a$.

Following the steps in Ref.~\cite{challinor1}, we expand the right-hand side of
Eq.~(\ref{e:temperature1}) in scalar harmonics and integrate along the line
of sight from the early universe to the observation point $R$. Neglecting
effects due to the finite thickness of the last scattering surface, on
integrating by parts we find that the temperature anisotropy involves
the quantity
\begin{equation}
\left(\frac{a}{k} \sigma_k' \right)' + \frac{1}{3} \frac{k}{a} (\sigma_k -
{\cal Z}_k) + A_k' - H A_k = - 2 \dot{\Phi}_k + \left(\frac{a}{k} \right)^2 I\;
\label{e:temperature10}
\end{equation}
integrated along the line of sight (after multiplying with $Q^{(k)}$).
In simplifying Eq.~(\ref{e:temperature10}) we have made use of the derivative
of the shear propagation equation
$\eqref{e:propagation2a}$, substituted for $q_k$ and
${\cal Z}_k$ from equations $\eqref{e:propagation1a}$ and
$\eqref{e:constraint1a}$, and finally used equations
$\eqref{e:friedmann1}$ and $\eqref{e:raychaudhuri}$.
The quantity $I$ is the total sum of all the braneworld corrections:
\begin{equation}
I  = \left(\frac{a}{k} \right)^2 \left[\dot{I}_1 + \frac{1}{3} \Theta I_1 + I_2
+ \frac{1}{3}  \left(\frac{k}{a} \sigma_k \right) I_3 +
\frac{1}{2} \left(\frac{k}{a} \right) I_4 \right]\;,
\end{equation}
where
\begin{equation}
\begin{split}
I_1 &= \frac{1}{24}\left(\frac{\tilde{\kappa}}{\kappa}\right)^4 [-(3 \g - 2)
\kappa^4
\rho^2 \pi_k + 12 \rho {\cal P}_k]\;,   \\
I_2 &= \frac{1}{72} \left(\frac{\tilde{\kappa}}{\kappa}\right)^4 \bigg\{
-\kappa^4 \bigg[6 \left(\frac{k}{a} \right) \g \rho^2 \sigma_k -
3(\dot{\rho} + 3 \dot{P}) \rho
\pi_k - 3 (3 \g -2) \rho (\rho \dot{\pi}_k + \dot{\rho} \pi_k) - 6
\left(\frac{k}{a} \right) \rho^2
q_k  \\
& - (3\g -2) \rho^2 \Theta \pi_k \bigg] - 48 \left(\frac{k}{a} \right) {\cal U}
\sigma_k - 36 (\dot{\rho} {\cal P}_k + \rho \dot{{\cal P}}_k) +
36 \left(\frac{k}{a} \right) \rho {\cal Q}_k - 12 \rho \Theta
{\cal P}_k \bigg\} \;, \\
I_3 &= \frac{1}{12} \left(\frac{\tilde{\kappa}}{\kappa}\right)^4 [(3 \g-1)
\kappa^4 \rho^2 + 12 {\cal U}]\;, \\
I_4 &= \frac{1}{18} \sigma_k \tilde{\kappa}^4 \rho^2
+ \frac{2}{3} \left(\frac{\tilde{\kappa}}{\kappa}\right)^4 {\cal U}
\sigma_k -  \frac{1}{24}\left(\frac{\tilde{\kappa}}{\kappa}\right)^4
(4 \kappa^4 \rho^2 q_k + 24 \rho {\cal Q}_k)\;.
\end{split}
\end{equation}
A lengthy calculation making use of the propagation and constraint equations
shows that $I=0$. The final result for the temperature anisotropies is then
\begin{equation} \label{e:temperature2}
\begin{split}
[\delta_{T}(e) ]_R &= -\sum_k \left[ \left( \frac{1}{4} \Delta_k^{(\g)} +
\frac{a}{k} \dot{\sigma}_k + A_k \right) Q^{(k)} \right]_A
+ \sum_k [(v_k^{(b)} - \sigma_k) e^a Q_a^{(k)} ]_A \\
& + \frac{3}{16} \sum_k (\pi_k^{(\g)} e^a e^b Q_{ab}^{(k)})_A
 + 2 \sum_k \int^{\lambda_R}_{\lambda_A}  \dot{\Phi}_k Q^{(k)} d\lambda\;,
\end{split}
\end{equation}
where the event $A$ is the intersection of the null geodesic with the last
scattering surface.
  
In retrospect, one could re-derive the result for the temperature anisotropy
in braneworld models much more simply by retaining the effective stress-energy
variables $\rho^{\text{tot}}$,
$P^{\text{tot}}$, $q_{a}^{\text{tot}}$ and $\pi_{ab}^{\text{tot}}$ in the
propagation and constraint equations used in the manipulation of the
left-hand side of Eq.~(\ref{e:temperature10}), rather than isolating the
braneworld contributions.

If we adopt the longitudinal gauge, defined by $\sigma_{ab}=0$, we find that
the electric part of the Weyl tensor and the acceleration are related by
$\Phi_k = -A_k$ if the total anisotropic stress $\pi_{ab}^{\text{tot}}$
vanishes. It follows that in this zero shear frame we recover the result
found by Langlois et al~\cite{langlois}.

Regarding the imprint of braneworld effects on the CMB, we note several
possible sources. Once the universe enters the low-energy regime the
dynamics of the perturbations are essentially general relativistic in the
absence of non-local anisotropic stress (see Sec.~\ref{sec:lowenergy}).
If ${\cal P}_{ab}$ really were zero, the only imprints of the braneworld on
the CMB could arise from modifications to the power spectrum (and
cross correlations) between the various low-energy modes. Since there are two
additional isocurvature modes in the low-energy universe due to braneworld
effects, it need not be the case that adiabatic fluctuations produced during
high-energy (single-field) inflation give rise to a low-energy universe
dominated by the growing, adiabatic, general-relativistic mode. The
possibility of exciting the low-energy isocurvature (brane) modes from
plausible fluctuations in the high-energy regime is worthy of further
investigation. In practice we do not expect ${\cal P}_{ab}=0$. In this case
the non-local anisotropic stress provides additional driving terms to the
dynamics of the fluctuations, and we can expect a significant manifestation
of five-dimensional Kaluza-Klein effects on the CMB anisotropies.

\section{Conclusion} \label{sec:discussion}

In this paper we have discussed the dynamics of cosmological scalar
perturbations in the braneworld scenario from the viewpoint of brane-bound
observers making use of the 1+3 covariant approach. We only considered
matter components present in the $\Lambda$CDM model, but it is straightforward
to include other components such as hot dark matter. 

We presented approximate, analytic solutions for the fluctuations in
the low-energy universe under the assumption that the non-local anisotropic
stress was negligible. We obtained two additional isocurvature modes not
present in general relativity in which the additional density gradients or
peculiar velocities of the total matter are compensated by fluctuations in
the non-local variables. In practice we do not expect the non-local anisotropic
stress ${\cal P}_{ab}$ necessarily to be negligible; in this case
our solutions to the homogeneous equations should form a useful starting point
for the construction of solutions to the driven equations. By adopting a
four-dimensional approach our presentation is necessarily limited. In
particular, we cannot predict the evolution of ${\cal P}_{ab}$
on the brane. However, the four-dimensional approach should be well-suited to
a phenomenological description of these five dimensional Kaluza-Klein modes.
A simple possibility is to adopt an ansatz for the evolution of
${\cal P}_{ab}$~\cite{barrow}, and this will be explored further in a future
paper.

We also presented solutions to the perturbation equations in the high-energy
regime where braneworld effects dominate. In this limit the gravitational
potential satisfies a fourth-order differential equation which we were unable
to solve analytically (even with ${\cal P}_{ab}=0$). Instead we constructed
power series solutions for the case where the non-local anisotropic stresses
vanish; these should prove useful for setting initial conditions in the
high-energy regime when performing a numerical solution of the perturbation
equations. We found two additional modes over those present in general
relativity, one of which can be described as a (brane) isocurvature velocity
mode. We also showed that the adiabatic decaying mode varies less rapidly
than in general relativity on large scales.

The detailed calculation of braneworld imprints on the CMB (in the
phenomenological approach discussed above) will be described in a future paper.
Here we showed with the 1+3 covariant approach that the line of sight
integral for the CMB temperature anisotropies is unchanged in form
from general relativity. We also noted that excitation of the additional
isocurvature modes present in the low-energy universe could provide an
additional imprint in the CMB, over and above that due to the non-local
anisotropic stress.

\begin{acknowledgments}

B.\ L., P.\ D., and A.\ L.\ thank the organisers of the
8th course of the International School of Astrophysics
``D.\ Chalonge'' held in Erice, 2000 at which this work was initiated.
B.\ L.\ thanks the Relativity and Cosmology
Group, University of Portsmouth for hospitality, and R.\
Maartens for useful correspondences over issues connected with
the non-local anisotropic stress, and for helpful comments on an earlier
draft of the paper. B.\ L.\ also thanks B.\ Bassett, C.\ Gordon, J.\ C.\
 Hwang, D.\ Langlois, A.\ Lewis, Y.\ L.\ Loh, C.\ J.\ A.\ P.\ Martins,
J.\ Soda, D.\ Wands and
C.\ Van de Bruck for insightful comments.
B.\ L.\ is supported by
an Overseas Research Studentship, the Cambridge Commonwealth Trust and
the Lee Foundation, Singapore.
P.\ D.\ thanks the NRF (South Africa) for financial support 
and the Cavendish Laboratory for hospitality. A.\ C.\ acknowledges
a PPARC Postdoctoral Fellowship.

\end{acknowledgments}

\appendix

\section{Energy frame equations in the radiation-dominated era}

In this appendix we present a complete set of evolution equations for the
total matter variables in the matter energy frame, $q_a = 0$. Note that
the four-velocity of the energy frame is not necessarily a timelike eigenvector
of the Einstein tensor in the presence of the non-local braneworld corrections
to the effective stress-energy tensor.
We assume that the matter is radiation dominated, the non-local
energy density vanishes in the background, and we ignore local anisotropic
stresses. We also assume that the baryons and CDM make a negligible
contribution to the fractional gradient in the total matter energy density
and to the energy flux, thus excluding the CDM and baryon isocurvature
modes. We also give the evolution equations for the non-local density
gradient and energy flux in the matter energy frame. 

Denoting the variables in the energy frame by an overbar\footnote{This
notation should not be confused with our use of the overbar to denote
rescaling by $\kappa^4 \rho$ in Sec.~\ref{power}.},
the relevant equations for scalar perturbations are
\begin{eqnarray}
\dot{\bar{\Delta}}_a &=&\frac{1}{3}\Theta\bar{\Delta}_a-\frac{4}{3}
\bar{{\cal Z}}_a\;, \\
{\dot{\bar{\cal Z}}}_a &=& -\frac{2}{3}\Theta \bar{\cal Z}_a -\frac{1}{4}
D^2 \bar{\Delta}_a
-\left(\frac{\tilde{\kappa}}{\kappa}\right)^4 \rho \bar{\Upsilon}_a
-\frac{1}{2} \kappa^2 \rho \bar{\Delta}_a
\left(1+\frac{5\rho}{\lambda}\right)\;,\\
\dot{\bar{\Upsilon}}_a &=& -\frac{a}{\rho} D^2 {\bar{\cal Q}}_a \;,\\
{\dot{\bar{\cal Q}}}_a &=& - \frac{4}{3}\Theta \bar{\cal Q}_a - \frac{\rho}{3a}
\bar{\Upsilon}_a - \frac{2 \kappa^4 \rho^2}{9 a}\bar{\Delta}_a -
D^b \bar{\cal P}_{ab}.
\end{eqnarray}

Solutions of these equations are related to those in the CDM frame
(Sec.~\ref{sec:radiation}) by linearising the frame transformations given in
Ref.~\cite{maartens4}. If the CDM projected velocity is $\bar{v}^{(c)}_a$
in the energy frame, the variables in the CDM frame are given by
\begin{eqnarray}
\Delta_a &=& \bar{\Delta}_a - \frac{4}{3} a \Theta \bar{v}^{(c)}_a\; ,\\
{\cal Z}_a &=& \bar{{\cal Z}}_a + \frac{1}{a} D_a D^b \bar{v}^{(c)}_b
- \frac{2\kappa^2 \rho}{a}\left(1+\frac{\rho}{\lambda}\right)\bar{v}^{(c)}_a
\;,\\
\Upsilon_a &=& \bar{\Upsilon}_a \; ,\\
{\cal Q}_a &=& \bar{{\cal Q}}_a \; , \\
q_a &=& - \frac{4}{3}\rho \bar{v}_a^{(c)}\; ,
\end{eqnarray}
where we have used ${\cal U}=0$ in the background.
The CDM peculiar velocity
evolves in the energy frame according to
\begin{equation}
\dot{\bar{v}}_a^{(c)} = - \frac{1}{3}\Theta \bar{v}^{(c)}_a + \frac{1}{4a}
\bar{\Delta}_a.
\end{equation}

\end{document}